\documentstyle[twocolumn,aps]{revtex}

\input{epsf}
\begin{document}

\def\aL{ a^{\dag}_{L} }
\def\aR{ a^{\dag}_{R} }
\def\bL{ b^{\dag}_{L} }
\def\bR{ b^{\dag}_{R} }

\draft
\title{Entangled quantum tunneling of two-component
Bose-Einstein condensates}
\author{H. T. Ng, C. K. Law, and P. T. Leung}
\address{{Department of Physics,
The Chinese University of Hong Kong,}\\ {Shatin, NT, Hong Kong,
China}}
\date{\today}
\maketitle

\begin{abstract}
We examine the quantum tunneling process in Bose condensates of
two interacting species trapped in a double well configuration. We
discover the condition under which particles of different species
can tunnel as pairs through the potential barrier between two
wells in opposition directions. This novel form of tunneling is
due to the interspecies interaction that eliminates the self-
trapping effect. The correlated motion of tunneling atoms leads to
the generation of quantum entanglement between two macroscopically
coherent systems.
\end{abstract}

\pacs{PACS numbers: 03.75.Fi, 03.65.Ud, 74.50.+r }

Quantum tunneling of macroscopically coherent systems is an
intriguing phenomena well known in the context of Josephson
junction effects in superconducting electronic systems. For
superfluids consisting of neutral particles, detailed
investigations of tunneling is aided by the recent realization
Bose-Einstein condensation of atomic vapor in a well controllable
environment. Indeed, recent experiments have successfully
demonstrated quantum tunneling for condensates confined in an
array of optical potentials \cite{kasevich1,burger}. One prominent
feature of tunneling in Bose condensates is the nonlinear dynamics
arising from the interaction between atoms. Quite remarkably, for
single-component condensates trapped in double-well
configurations, previous studies have indicated that a
self-trapping mechanism can suppress the tunneling rate
significantly by increasing the atom-atom interaction strength
\cite{shenoy,walls,raghavan,williams}.

An interesting extension of the tunneling problem involves Bose
condensates of two interacting species (Fig. 1). The main issue is
how the interspecies interaction affects the tunneling process,
and particularly the quantum coherence as the two condensates mix
together. Previous studies of the general properties of
two-component Bose condensates have emphasized the important role
of the interspecies interaction, which leads to novel features,
such as the components separation \cite{phase1,phase2},
cancellation of the mean field energy shift \cite{pu}, and the
suppression of quantum phase diffusion \cite{law}. However, the
investigation of the influence of interspecies interaction on
tunneling dynamics has only just begun \cite {lobo,pu2}.

In this paper we present a novel tunneling mechanism for a
two-component condensate trapped in a double-well (see Fig. 1).
The atoms of the component $A(B)$ are initially prepared in the
left (right) potential well. We discover the condition under which
the interspecies interaction can eliminate the self-trapping
effect and thus enhances the tunneling significantly. Such an
enhanced tunneling originates from the correlated quantized motion
of the two condensates. We also show that atoms of different
species tunnel through the barrier as correlated pairs in
opposition directions, i.e., a form of {\em quantum entangled
tunneling}. Therefore tunneling serves as a mechanism to build up
a strong correlation among atoms of different species, and this
leads to the generation of quantum entanglement between two
multi-particle systems.

The configuration of our double-well system is sketched in Fig. 1.
Our focus in this paper is the quantum dynamics beyond the mean
field description. An exact many-body description is difficult
even for single-component condensate problems. The usual method to
capture the essential physics is based on the two-mode
approximation in which the evolution is confined by the left and
right localized mode functions associated with the respective
potential wells \cite{shenoy,walls,raghavan,williams,juha}. Such
an approximation is valid when each potential well is sufficiently
deep so that higher modes of the wells essentially do not
participate in the dynamics.

\begin{figure}
\vspace{-5mm} \centerline{\epsfxsize=2.5in \epsfbox{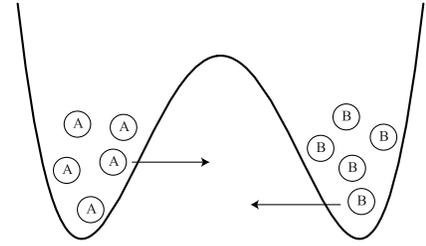}}
\vspace{5mm} \caption{A sketch of the double-well tunneling
system. Initially, the atoms of the component $A$ are trapped in
the left side and the atoms of the component $B$ are trapped in
the right side.}
\end{figure}

In the two-mode approximation, the system is modeled by
the Hamiltonian $(\hbar =1)$,
\begin{eqnarray}
H &=& \frac{\Omega}{2}
({\aL}a_{R}+{\aR}a_{L}+{\bL}b_{R}+{\bR}b_{L}) \nonumber \\ &&
+\frac{\kappa_{a}}{2}\left[
({\aL}a_{L})^{2}+({\aR}a_{R})^{2}\right] \nonumber \\ && +
\frac{\kappa_{b}}{2}\left[({\bL}b_{L})^{2}+({\bR}b_{R})^{2}\right]
\nonumber \\ && +\kappa
({\aL}a_{L}{\bL}b_{L}+{\aR}a_{R}{\bR}b_{R}). \label{Hamiltonian}
\end{eqnarray}
Here the subscripts $L$ and $R$ respectively denote the localized
modes in the left and right potential wells. Since there are two
modes available for each component, the model in fact consists of
four mode operators. We use $a^{\dag}_j$ and $b^{\dag}_j$
$(j=L,R)$ to denote the creation operators of the component $A$
and $B$ respectively. The parameters $\Omega$,
$\kappa_a(\kappa_b)$ and $\kappa$ describe the tunneling rate,
self-interaction strength of the component $A(B)$ and the
interspecies interaction strength respectively.

To gain insight of the quantum correlation developing in the
tunneling process, we first consider the exactly solvable case
with only one $A$ atom in the left well and one $B$ atom in the
right well. In this case the system is spanned by four basis
vectors: $|1,0\rangle_{A}|1,0\rangle_{B}$,
$|1,0\rangle_{A}|0,1\rangle_{B}$, $|0,1\rangle_{A}|1,0\rangle_{B}$
and $|0,1\rangle_{A}|0,1\rangle_{B}$, where $|p,q\rangle_{S}$
denotes the state with $p$ atoms of species $S$ $(S=A,B)$ in the
left well and $q$ atoms of species $S$ in the right well. The
eigenvalues and eigenvectors of $H$ can be found
straightforwardly. In the regime where the interspecies
interaction is sufficiently strong such that $\kappa \gg \Omega$,
the state vector evolves as
\begin{eqnarray}
|\Psi(t)\rangle &=& e^{-i[(\kappa_{a}+\kappa_{b})/2-\omega_0]{t}}
[\cos{\omega_{0}{t}}|1,0\rangle_{A}|0,1\rangle_{B} \nonumber \\ &&
+ i\sin{\omega_{0}{t}}|0,1\rangle_{A}|1,0\rangle_{B}] + O (\Omega
/ \kappa) . \label{2atom-state}
\end{eqnarray}
In writing Eq. (\ref{2atom-state}) we have defined $\omega_0 =
\Omega^2/2 \kappa$ as an effective tunneling frequency. Because of
the strong interaction between the atoms, the probability of
finding both particles in the same well at any time $t$ is
negligible (of order $\Omega^2/\kappa^2$). The tunneling motion of
the two atoms are anti-correlated in the sense that the atom $A$
and the atom $B$ always move in opposite directions. Such an
anti-correlated tunneling motion gives rise to quantum
entanglement between the two atoms. At time $t=(n+1/4)\pi /
\omega_0$, ($n=$ integers), the state is a form of Bell's state
that is maximally entangled in the two-particle two-mode subspace.

Now we examine the multiple atoms case. In order to facilitate the
discussion, we assume the number of particles are the same for the
two components, i.e., $N_a=N_b=N$, and the condensates shares the
same interaction strength i.e., $\kappa_a=\kappa_b=\kappa$. The
latter condition is a good approximation to  $^{87}${Rb}
condensate of atoms in hyperfine spins states
$|F=2,m_{f}=1{\rangle}$ and $|F=1,m_{f}=-1{\rangle}$, which share
similar scattering lengths \cite{phase2}. However, we emphasize
that these assumptions are not crucial, we shall relax these
conditions later in the paper.

We shall limit our study to the $4 \kappa {\gg} N\Omega$ regime
where the nonlinear interaction is dominant. As before we consider
the initial condition in which all atoms in the component $A(B)$
are localized in the left (right) potential well. The general form
of the state vector at time $t$ is given by: $|{\Psi}(t){\rangle}=
e^{-i{\kappa}N^{2}t} \sum_{n=0}^{N}\sum_{m=0}^{N} {c}_{n,m}(t)
|n,N-n{\rangle}_A|m,N-m{\rangle}_B$. The amplitudes ${c}_{n,m}
(t)$ are governed by the Schr\"odinger equation according to the
Hamiltonian (\ref{Hamiltonian}):
\begin{eqnarray}
\label{qpamp} i\dot{{c}}_{n,m}&=&\frac{\Omega}{2} \left[
\sqrt{(n+1)(N-n)}{{c}}_{n+1,m} \right. \nonumber \\ && \left.  \ \
\ \ \ \ + \sqrt{n(N-n+1)}{{c}}_{n-1,m}\right] \nonumber\\ && +
\frac{\Omega}{2}\left[\sqrt{(m+1)(N-m)}{{c}}_{n,m+1} \right.
\nonumber \\ && \left. \ \ \ \ \ \ +
\sqrt{m(N-m+1)}{{c}}_{n,m-1}\right] \nonumber
\\ && + {\kappa}\left[(n+m-N)^{2}\right]{{c}}_{n,m}
\end{eqnarray}
with the initial condition $c_{n,m} (0)= \delta_{n,N}
\delta_{m,0}$.

Let us first present the typical results obtained from the
numerical solutions of Eq.~(3). In Fig. 2, we show the particle
number difference $W \equiv \langle {a_L^{\dag} a_L  - a_R^{\dag}
a_R } \rangle$ of species $A$ between the two wells as a function
of time. The occurrence of tunneling is revealed by the decrease
of $W$. At longer times $W$ approaches zero, therefore the numbers
of $A$ atoms in the two potential wells are roughly equalized. We
emphasize that the nonzero interspecies interaction is responsible
for the tunneling to occur. If the two species do not interact
with each other (i.e., $\kappa =0$), then a sufficiently strong
self-interaction $\kappa_j> N \Omega$ $(j=a,b)$ can suppress the
tunneling almost completely by the self-trapping effect
\cite{shenoy}.

\begin{figure}
\centerline{\epsfxsize=2.8in \epsfbox{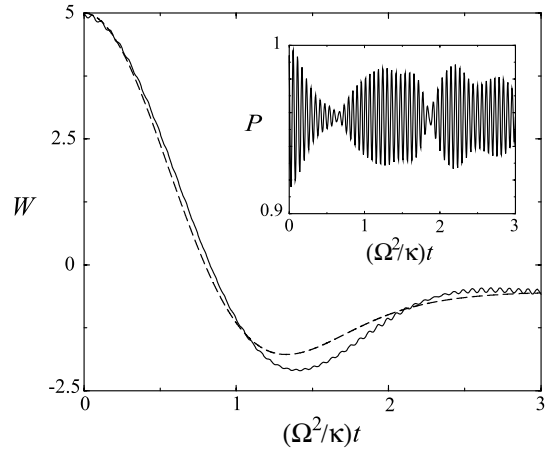}} \vspace{5mm}
\caption{Particle number difference of the component $A$ between
the two potential wells as a function of dimensionless time
$(\Omega^2/\kappa){t}$ with $N=5$ and $\kappa=\kappa_a= \kappa_b=
10 \Omega$. The dashed line is the solution based on the effective
Hamiltonian in Eq. (6). The inset shows the overlap probability
$P$ of finding the system belonging to the degenerate set of
states $|\phi_q \rangle$. }
\end{figure}

To understand the mechanism of the two-component tunneling, we
note that the state vector (3) is a superposition of $(N+1)^2$
states of the form $|n ,N-n {\rangle}_A|m,N-m{\rangle}_B$.
However, only a few number of them are actively involved. This is
intuitively understood because $|n ,N-n {\rangle}_A
|m,N-m{\rangle}_B$ have different energies for different values of
$n$ and $m$, and only those with energies closer to that of the
initial state are more accessible.

The difference in energies is significant in the regime $4 \kappa
\gg N \Omega$ considered here. We find that the states
\begin{equation}
|\phi_q \rangle =|q ,N-q {\rangle}_A |N-q,q {\rangle}_B \ \ \ \
(q=0,1,2,...,N) \label{degenerate-state}
\end{equation}
are approximately degenerate energy eigenvectors of $H$ (to zero
order in $\Omega$), and the energies of states other than $|\phi_q
\rangle$ are higher than that of $|\phi_q
\rangle$ by an amount of order of $\kappa$ or higher. Since the
initial state $|\Psi(0){\rangle}$ is $|\phi_0 \rangle$,
$|\Psi(t){\rangle}$ is mainly a superposition of $|\phi_q \rangle$
at any time $t$ according to the energy argument above. We find
that this is indeed the case. To provide a numerical evidence of
our finding, we show in the inset of Fig. 2 the overlap
probability defined by $P(t) = \sum\limits_{q = 0}^N {}
|\left\langle {{\phi _q }}
 \mathrel{\left | {\vphantom {{\phi _q } \Psi }}
 \right. \kern-\nulldelimiterspace}
 {\Psi (t) } \right\rangle |^2$.
For the parameters used in this figure, the set of $|\phi_q
\rangle$ contributes more than 90$\%$ of $|\Psi(t){\rangle}$. Our
further numerical tests suggest that $P \to 1$ in the limit
$N\Omega / \kappa \to 0$.

We remark that all $|\phi_q \rangle$ have the same number of
particles (component $A$ plus component $B$) in the left potential
well, and the same also holds for the right well. Therefore $P(t)
\approx 1$ implies small fluctuations of total particle number in
the each potential well. In the case of Fig. 2, we find that the
fluctuation of total particle number in the left well
$\left\langle {\Delta N_L } \right\rangle$ is much smaller than $
\left\langle {N_L } \right\rangle ^{1/2}$, i.e., a sub-Poissonian
distribution.

The time-dependent problem now is significantly simplified because
the evolution of the condensates mainly involves the set of
degenerate states $|\phi_q \rangle$. For those states that do not
have the same energy as $|\phi_q \rangle$, they act as
intermediate states that are rarely populated. We may eliminate
such intermediate states with the adiabatic following method which
is well known in quantum optics. The last line in Eq. (3)
indicates that the states with $n+m=N+r$ ($r=$integer) has the
diagonal term $r^2\kappa$ that can be interpreted as the energy of
the states if the tunneling interaction $\Omega$ were switched
off. Since $\kappa$ is a large parameter here, the states with
$n+m=N \pm 1$ have a much different energy than that of the states
$|\phi_q \rangle$ with $n+m=N$. Any transition (due to $\Omega$)
from the manifold $n+m=N$ to the manifolds $n+m=N \pm 1$ must
quickly return to $n+m=N$. In other words, the states with $n+m=N
\pm 1$ are intermediate states that are hardly occupied at any
time. The amplitudes associated with the $n+m=N \pm 1$ manifold
are: ${c}_{n+1,N-n}$, ${c}_{n-1,N-n}$, ${c}_{n,N-n+1}$ and
${c}_{n,N-n-1}$, and they can be found approximately by adiabatic
following under the condition $4\kappa{\gg}{N}\Omega$. With these
approximate amplitudes, the Schr\"{o}dinger equation of
${c}_{n,N-n}$ is given by:
\begin{eqnarray}
i\dot{{c}}_{n,N-n}&{\approx}&-\frac{\Omega^{2}}{2\kappa}
(n+1)(N-n){c}_{n+1,N-n-1} \nonumber \\ && -
\frac{\Omega^{2}}{2\kappa}n(N-n+1) {c}_{n-1,N-n+1} \nonumber\\
&&-\frac{\Omega^{2}}{\kappa}[n(N-n)+N] {c}_{n,N-n}.
\label{adabatic-eqn}
\end{eqnarray}
This corresponds to the Schr\"{o}dinger equation governed by the
effective Hamiltonian:
\begin{eqnarray}
H_{\rm{eff}}&=&-\frac{\Omega^{2}}{2\kappa} \left( {\aL}a_{R}b_{L}
{\bR}+a_{L}{\aR}{\bL}b_{R}+{\aL}a_{L}{\aR}a_{R} \right. \nonumber
\\ && \left. \ \ \ \ \ \ \ \ +{\bL}b_{L}{\bR}b_{R} \right),
\label{effective-Hamiltonian1}
\end{eqnarray}
apart from a constant term proportional to the total number of
particles. The effective Hamiltonian $H_{\rm{eff}}$ is an
approximation that captures the essential tunneling mechanism in
the $4 \kappa \gg N \Omega$ limit. We have tested the validity of
the effective Hamiltonian numerically. For example, in Fig. 2 the
dashed line, which is obtained from the evolution based on
$H_{\rm{eff}}$, agrees well with the exact numerical solution.

The physical picture becomes transparent according to
$H_{\rm{eff}}$. The interaction term ${\aR}{\bL}a_{L}b_{R}$
suggests that every time when an atom $A$ moves from the left well
to the right well there must be an atom $B$ moves from the right
well to the left well. This explains why the small fluctuations in
the total particle number in each potential well. The reverse
process is described by ${\aL}{\bR}a_{R}b_{L}$ in $H_{\rm{eff}}$.
In other words, the atoms $A$ and $B$ must be moving in pair in
opposite directions during the tunneling  process.

It is worth noting that $H_{\rm{eff}}$ can be cast in the form
\begin{equation}
H_{\rm{eff}}=-\frac{\Omega^{2}}{4\kappa}\left( {
K_{+}K_{-}+K_{-}K_{+}} \right) \label{effective-Hamiltonian2}
\end{equation}
where $K_{+}= a^{\dag}_{L}a_{R}+ b^{\dag}_{L}b_{R} $ and $K_{-}=
a_{L}a^{\dag}_{R}+ b_{L}b^{\dag}_{R}$ satisfy the angular momentum
commutation relation: $ \left[ {K_ +  ,K_ -  } \right] = 2K_z $, $
\left[ {K_z ,K_ \pm  } \right] =  \pm K_ \pm$, where $K_z  =
(a_L^{\dag} a_L  + b_L^{\dag} b_L  - a_R^{\dag} a_R - b_R^{\dag}
b_R)/2$. Therefore $K_{\pm}$ and $K_z$ are analogous to collective
spin operators, and the eigenvectors and eigenvalues of
$H_{\rm{eff}}$ can be solved analytically using angular momentum
algebra. This shares a common feature in spinor condensates where
the spin mixing problem can be solved in a similar fashion
\cite{spin}. The nonlinear interaction between collective spins is
the key to generate nonclassical correlations between different
spin components, and particularly, the notion of multi-particle
entanglement has been discussed in the context of squeezed spin
states \cite{duan,multi,you}. We show here that quantum
entanglement between two multi-particle subsystems ($A$ and $B$)
can be achieved in the double-well tunneling process.

To this end we quantify the degree of  entanglement between the
two species based on entanglement entropy:
$E=-\mbox{tr}(\rho_{A}\mbox{ln}\rho_{A})=-\mbox{tr}(\rho_{B}
\mbox{ln}\rho_{B})$, where $\rho_{A}$ and $\rho_{B}$ are reduced
density matrices of the respective subsystems, i.e.,
$\rho_{A}=\mbox{tr}_{B}{\rho_{AB}}$ and
$\rho_{B}=\mbox{tr}_{A}{\rho_{AB}}$ with $ \rho_{AB} =\left| {\Psi
(t)} \right\rangle \left\langle {\Psi (t)} \right|$ being the
density matrix of the whole system. A disentangled state (for
example the initial state above) has zero entanglement entropy.
The more entangled the systems, the larger the value of $E$ is. As
an illustration, we show in Fig. 3 how the entanglement is
established in time for various particle numbers. As the time
increases the value of $E$ increases until a saturated value is
reached. Since there are $N+1$ degenerate states $|\phi_q \rangle$
mainly involved in the evolution, $E\approx \ln{(N+1)}$ is a
maximum if all $|\phi_q \rangle$ have equal contributions to the
state of the system. In the case of Fig. 3, the value of $E$ can
reach as high as 90\% of the value $\ln (N+1)$.

Finally we would like to address the conditions for the entangled
tunneling to occur. Our discussion above has been restricted to
the simplest symmetric situation: $N_a=N_b=N$ and
$\kappa_a=\kappa_b=\kappa$, in order to illustrate the essential
mechanism under the condition $4\kappa \gg N \Omega$. The same
analysis can be performed to examine the general situations. We
have examined the system with unequal particle numbers $(N_b-N_a)
\equiv D >1$ and unequal coupling strengths
$\kappa_a=\kappa+\delta$, $\kappa_b=\kappa-\delta$ with $|\delta
|\ll \kappa$. We find that if the tunneling strength $\Omega$ is
sufficiently weak or the self interaction strength $\kappa$ is
sufficiently strong such that
\begin{equation}
4|\kappa (D-1)-[N_a(D-2)|\delta|]/2 | \gg \Omega N_b.
\label{general-condition}
\end{equation}
then the system mainly evolves among the states $|n ,N_a-n
{\rangle}_A |N_a-n,N_b-N_a+n{\rangle}_B$, where $n=0,1,2,...N_a$.
In other words, the tunneling under condition (8) is characterized
by entanglement generation term ${\aR}{\bL}a_{L}b_{R}$ as before.
However, we point out that the tunneling are generally less
efficient for the cases with non-zero $\delta$ and $D$. This is
because the states $|n ,N_a-n {\rangle}_A
|N_a-n,N_b-N_a+n{\rangle}_B$ are not as degenerate as that in the
symmetric case with $\delta$ and $D$ are both zero. We also remark
that the condition (8) does not apply to the special case $N_b-N_a
= \pm 1$ in which some of the states in the $n+m=N$, and $n+m=N
\pm 1$ manifolds are accidentally degenerated.

\begin{figure}
\centerline{\epsfxsize=2.8in \epsfbox{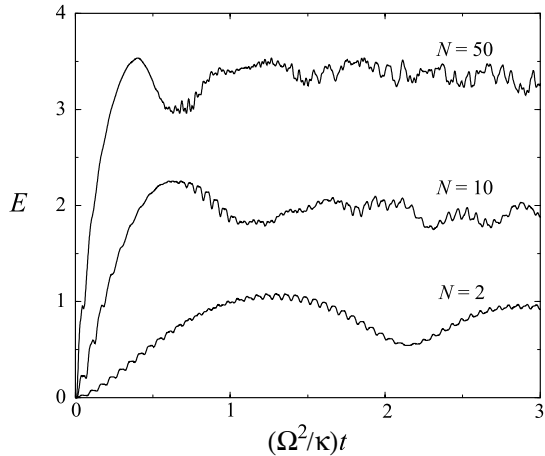}} \vspace{6mm}
\caption{The time dependence of the entanglement entropy $E$ for
the different particle numbers: $N=2$, $N=10$ and $N=50$ with
$\kappa=\kappa_a= \kappa_b= 10 \Omega$. }
\end{figure}

To conclude, we have presented a novel mechanism of double-well
tunneling involving Bose condensates of two interacting
components, based on the two-mode approximation model in the
strong coupling regime.  We find that the interplay of
intraspecies and interspecies interactions permits a set of energy
degenerate states, a small tunneling coupling can push the system
to `explore' through these degenerate states and thus result in a
substantial tunneling not limited by the self trapping effect. The
most interesting feature is the strongly correlated tunneling
motion. We have derived an exactly solvable effective Hamiltonian
to capture the mixing dynamics. In addition, we have shown that
high degree of quantum entanglement between two macroscopically
coherent systems can be achieved.

\acknowledgments We acknowledge the support from an RGC Earmarked
Grant CUHK4237/01P and the Chinese University of Hong Kong Direct
Grant (Grant Nos: 2060148 and 2060150).


\begin{references}



\bibitem{kasevich1} B.P. Anderson and M.A. Kasevich, Science
{\bf 282}, 1686 (1998).

\bibitem{burger} F.S. Cataliotti {\it et al.}, Science {\bf 293},
843 (2001).

\bibitem{shenoy} A. Smerzi, S. Fantoni,. S. Giovanazzi,
and S.R. Shenoy, Phys. Rev. Lett. {\bf 79}, 4950 (1997).

\bibitem{walls}
G.J. Milburn, J. Corney , E.M. Wright and D.F. Walls, Phys. Rev. A
{\bf 55}, 4318 (1997).

\bibitem{raghavan}
S. Raghavan, A. Smerzi, S. Fantoni, and S.R. Shenoy, Phys. Rev. A
{\bf 59}, 620 (1999).

\bibitem{williams}
J.E. Williams, Phys. Rev. A {\bf 64}, 013610 (2001).

\bibitem{phase1} T.-L. Ho and V.B. Shenoy, Phys. Rev. Lett.
{\bf 77}, 3276 (1996); H. Pu and N.P. Bigelow, Phys. Rev. Lett.
{\bf 80}, 1130 (1998); E. Timmermans, Phys. Rev. Lett. {\bf 81},
5718 (1998).

\bibitem{phase2} D.S. Hall {\it et al.},
Phys. Rev. Lett. {\bf 81}, 1539 (1998).

\bibitem{pu}
E.V. Goldstein, M.G. Moore, H. Pu, and P. Meystre, Phys. Rev.
Lett. {\bf 85}, 5030 (2000).

\bibitem{law}
C.K. Law, H. Pu, N.P. Bigelow, and J.H. Eberly, Phys. Rev. A {\bf
58}, 531 (1998).


\bibitem{lobo} S. Ashhab and C. Lobo,
Phys. Rev. A {\bf 66}, 013609 (2002).


\bibitem{pu2} H. Pu, W. Zhang, and P. Meystre,
Phys. Rev. Lett. {\bf 89}, 090401 (2002).


\bibitem{juha} J. Javanainen and M.Y. Ivanov, Phys. Rev. A
{\bf 60}, 2351 (1999).


\bibitem{spin}C.K. Law, H. Pu, and N.P. Bigelow,
Phys. Rev. Lett. {\bf 81}, 5257 (1998); E.V. Goldstein and P.
Meystre, Phys. Rev. A {\bf 59}, 3896 (1999).


\bibitem{duan}L.-M. Duan, J.I. Cirac, and P. Zoller,
Phys. Rev. A {\bf 65}, 033619 (2002).


\bibitem{multi}
A. Sorensen, L.M. Duan, I. Cirac and P. Zoller, Nature(London)
{\bf 409}, 63 (2001).

\bibitem{you} K. Helmerson and L. You, Phys. Rev. Lett. {\bf 87},
170402 (2001).

\end{references}
\end{document}